\documentclass[conference]{IEEEtran}
\IEEEoverridecommandlockouts

\usepackage{threeparttable}
\usepackage{amsthm}
\usepackage{textcomp}
\usepackage{booktabs}
\usepackage{bm}
\usepackage{multirow}
\usepackage{algorithm}
\usepackage{epstopdf}
\usepackage{epsfig}
\usepackage{enumitem}
\usepackage{booktabs}
\usepackage{diagbox}
\usepackage{amsmath,amsthm}

\usepackage{multirow}
\usepackage[normalem]{ulem}
\usepackage{makecell}
\usepackage[pscoord]{eso-pic}
\usepackage{xcolor}
\usepackage{stfloats}
\usepackage{titlesec}
\newtheorem{Definition}{Definition}

\newcommand{\qdel}[1]{}

\newcommand{\Qdel}[1]{}

\newcommand{\tdel}[1]{}

\newcommand{\hdel}[1]{}

\newcommand{\tabincell}[2]{\begin{tabular}{@{}#1@{}}#2\end{tabular}}

\def\BibTeX{{\rm B\kern-.05em{\sc i\kern-.025em b}\kern-.08em
    T\kern-.1667em\lower.7ex\hbox{E}\kern-.125emX}}
\begin{document}

\title{PrefixLLM: LLM-aided Prefix Circuit Design
}

\author{%
{\normalsize Weihua Xiao, Venkata Sai Charan Putrevu, Raghu Vamshi Hemadri, Siddharth Garg, Ramesh Karri}\\
{\normalsize New York University, New York, USA}\\
{\normalsize Emails: \{wx2356, v.putrevu, rh3884\}@nyu.edu, siddharth.j.garg@gmail.com, rkarri@nyu.edu}
}

\maketitle
\begin{abstract}
Prefix circuits are fundamental components in digital adders, widely used in digital systems due to their efficiency in calculating carry signals. 
Synthesizing prefix circuits with minimized area and delay is crucial for enhancing the performance of modern computing systems. 
Recently, \textit{large language models} (\textit{LLMs}) have demonstrated a surprising ability to perform text generation tasks. 
We propose \textit{PrefixLLM}, that leverages LLMs for prefix circuit synthesis. 
\textit{PrefixLLM} transforms the prefix circuit synthesis task into a structured text generation problem, termed the \textit{Structured Prefix Circuit Representation} (\textit{SPCR}), and introduces an iterative framework to automatically and accurately generate valid SPCRs. 
We further present a \textit{design space exploration} (\textit{DSE}) framework that uses LLMs to iteratively search for area and delay optimized prefix circuits. 
Compared to state-of-the-art, PrefixLLM can reduce the area by $3.70\%$ under the same delay constraint. 
This work highlights the use of LLMs  in the synthesis of arithmetic circuits, which can be transformed into the structured text generation.


\end{abstract}

\begin{IEEEkeywords}
Prefix Circuit, Large Language Model, Structured Text Generation, Design Space Exploration
\end{IEEEkeywords}
\section{Introduction}
\label{sec:intro}
Computing carry signals efficiently is  the most challenging aspect of designing digital adders~\cite{Brent82}. 
In an adder, the carry signal must be computed for each bit position to compute the final sum result.  For digital adders with large bit-widths, the delay in computing each carry signal is a bottleneck to the speed of the adder~\cite{Hauck00}.
Prefix circuits are an effective solution to this challenge~\cite{Kogge73,Brent82}. 
Instead of computing the carry signal bit-by-bit, prefix circuits convert the carry signal computation at each bit position into computing \textit{generate} and \textit{propagate} signals over multiple bits for each bit position~\cite{Brent82,Xiao21}. 
Prefix circuits take advantage of the basic operator \textit{prefix node} to compute generate and propagate signals in parallel, reducing the overall delay. 
Prefix circuits are classified based on their topological configurations, each providing different trade-offs between area and delay.
Several classical topological configurations, such as Sklansky~\cite{Sklansky60}, Kogge-Stone~\cite{Kogge73}, and Brent-Kung~\cite{Brent82}, are designed to balance the area and delay of prefix circuits, optimizing performance and resource utilization for specific constraints.
In~\cite{Snir86}, it proposes a theoretical lower bound on the area of a prefix circuit for a given delay.
Synthesizing a prefix circuit with the optimal trade-off between area and delay is still an open question due to the vast design space of topological configurations of a prefix circuit.

Various methods have been proposed for optimizing the trade-off between area and delay of a prefix circuit, with each approach offering unique advantages and limitations. 
These methods can be categorized into: classical prefix circuits, dynamic programming-based methods, and machine learning-based methods.
Classical prefix circuits, such as Sklansky~\cite{Sklansky60}, Kogge-Stone~\cite{Kogge73}, and Brent-Kung~\cite{Brent82}, feature highly regular structures that facilitate efficient carry signal computation. While these designs are straightforward and well-suited for general applications, they lack the flexibility to adapt to specific design constraints, such as unique area or delay requirements, limiting their applicability in specialized scenarios.
The dynamic programming-based methods model prefix circuit synthesis as a divide-and-conquer problem, which can be addressed using dynamic programming~\cite{Liu03,Roy13,Lin24}. 
By recursively dividing the problem into smaller sub-problems, dynamic programming explores the design space of prefix circuits systematically. 
However, the vast design space can result in low efficiency and high memory consumption of dynamic programming. 
Heuristics are used to prune the design space while preserving design quality. 
Despite these advances, designing effective heuristics that achieve a good trade-off between exploration efficiency and design quality is challenging.

Machine learning-based approaches model prefix circuit synthesis as a Markov Decision Process (MDP), which is then solved using \textit{reinforcement learning} (\textit{RL}) algorithms~\cite{Roy21,Lai24}. 
For example, \textit{PrefixRL}~\cite{Roy21} uses \textit{Deep Q Network} (\textit{DQN}) algorithm to optimize the prefix circuit synthesis policy, while the work in~\cite{Lai24} applies \textit{Monte-Carlo Tree Search} (\textit{MCTS}) algorithm.
However, they require careful design of the state space and action space, as these significantly influence the policy training. 
This dependence on expert knowledge in RL is a significant barrier to widespread adoption and limits the generalizability of these RL-based approaches.

We propose a novel LLM-aided framework \textit{PrefixLLM} that leverages large language models (LLMs) for synthesizing optimized prefix circuits without designing any heuristic or extra training. 
The main contributions are:
\begin{enumerate}
[noitemsep,nolistsep,leftmargin=*]
    \item [(1)] We for the first time apply LLM to synthesize prefix circuits, which can achieves results that are comparable to state-of-the-art techniques in terms of area and delay;
    \item [(2)] We transform the prefix circuit synthesis problem into a text generation task with a specific format, referred to as the \textit{Structured Prefix Circuit Representation (SPCR)}, which aligns well with the strengths of LLMs.
    \item [(3)] We  propose a LLM-aided DSE framework to  automatically search for prefix circuits with better area and delay.
    \item [(4)]  PrefixLLM framework can  synthesize other types of arithmetic circuits.
\end{enumerate}




In the remainder of this paper, Section~\ref{sec:pre} introduces some preliminaries. Section~\ref{subsec:valid_prefix_circuit} introduces the LLM-aided framework for synthesizing optimized valid prefix circuits. Then, the experiment results are reported in Section~\ref{sec:exp}. Finally, Section~\ref{sec:con} concludes the paper.

\section{Preliminaries} \label{sec:pre}

\subsection{Prefix circuit}
\label{subsec:PrefixCircuit}
A prefix circuit is a specialized type of digital circuit that can be used for generating carry signals in parallel, making it an essential component in digital adders~\cite{Brent82,Harris03}. 
Thus, the prefix circuit plays a critical role in modern computational hardware, particularly in high-speed processors, where efficient addition of binary numbers is crucial.

\begin{figure}
    \centering
    \includegraphics[width=\linewidth]{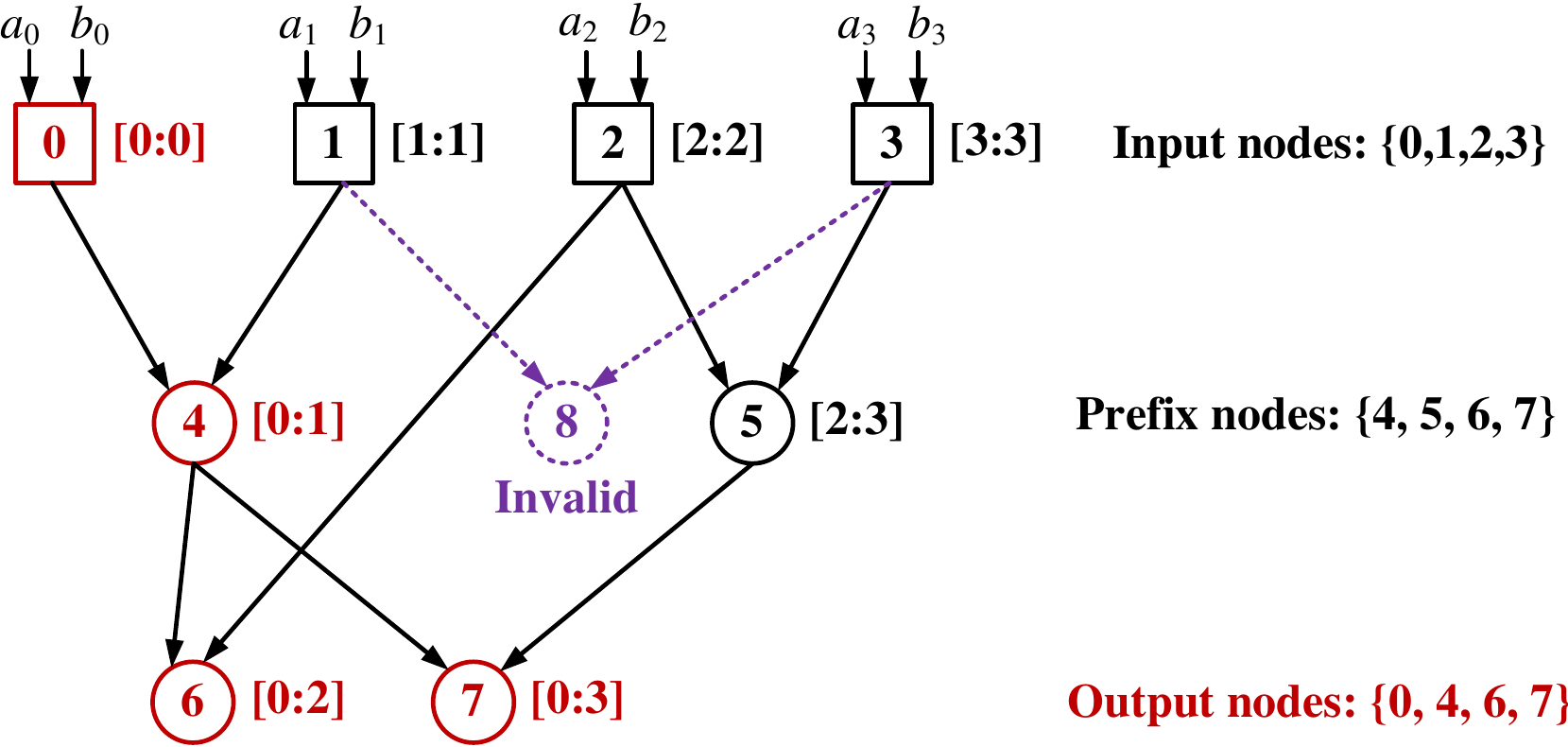}
    \caption{A $4$-bit valid prefix circuit with $4$ input nodes (nodes $0$-$3$) and $4$ valid prefix nodes (nodes $4$-$7$). Node $8$ is an example invalid prefix node.}
    \label{fig:preliminary-4-bit-prefix-circuit}
\end{figure}

In an $n$-bit digital adder with two operands $A=a_{n-1}a_{n-2}\cdots a_{0}$ and $B=b_{n-1}b_{n-2}\cdots b_{0}$, an $n$-bit prefix circuit, which can compute each carry signal $c_{i}$ $(0\leq i \leq n-1)$, is denoted as \textit{Valid Prefix Circuit}.
A valid prefix circuit is constructed using two types of basic operators: \textit{input node} and \textit{prefix node}, which are both to compute \textit{propagate} and \textit{generate} signals.
For an $n$-bit prefix circuit, it contains $n$ input nodes. 
The inputs of the $i$-th $(0\leq i \leq n-1)$ input node are the two bits $(a_{i},b_{i})$ from two operands $A$, $B$ at each bit position.
The $i$-th $(0\leq i \leq n-1)$ input node is used to compute the propagate and generate signals for the single $i$-th bit.
They are also generalized as the propagate and generate signals for the bit range $[i:i]$, denoted as $p_{i:i}$ and $g_{i:i}$ respectively, where:
\begin{equation}
\label{eqn:input_node}
    p_{i:i}=a_{i}\oplus b_{i},\ g_{i:i}=a_{i}b_{i}
\end{equation}
For simplicity, we denote the bit range, which the propagate and generate signals of a node are computed for, as the bit range of the node.
Fig.~\ref{fig:preliminary-4-bit-prefix-circuit} illustrates an example of a $4$-bit valid prefix circuit, featuring $4$ input nodes (nodes $0$-$3$) represented as squares, with the corresponding bit range displayed next to each input node.
For each prefix node, it combines two groups of propagate and generate signals of two predecessor nodes for smaller ranges of bits, denoted as $(p_{i:j},g_{i:j})$ and $(p_{k:l},g_{k:l})$ where $i\leq j \leq k \leq l$, to compute the propagate and generate signals for a larger range of bits.
The formulations of the propagate and generate signals are shown in Eqs.~\ref{eqn:propagate_signal}, \ref{eqn:generate_signal}.
\begin{equation}
\label{eqn:propagate_signal}
p_{i:j}p_{k:l}
\end{equation}
\begin{equation}
\label{eqn:generate_signal}
g_{i:j}+p_{i:j}g_{k:l}
\end{equation}
However, a prefix node in a valid prefix circuit can not combine any two groups of propagate and generate signals.
A valid prefix circuit can be constructed only using \textit{Valid Prefix Nodes}, which satisfy the condition in Eq.~\ref{eqn:condition_prefix_node}. 
\begin{equation}
\label{eqn:condition_prefix_node}
k=j+1
\end{equation}
Then, the  bit range of the prefix node is $[i:l]$ and the computed propagate and generate signals are $(p_{i:l},g_{i:l})$.
In Fig.~\ref{fig:preliminary-4-bit-prefix-circuit}, the valid prefix circuit uses $4$ valid prefix nodes (nodes $4$-$7$), denoted as circles.
For example, node $5$ combines two groups of propagate and generate signals for bit ranges $[2:2]$ and $[3:3]$, which satisfy Eq.~\ref{eqn:condition_prefix_node}, and the resulting bit range is $[2:3]$. Fig.~\ref{fig:preliminary-4-bit-prefix-circuit},  shows an example invalid prefix node $8$.

According to~\cite{Dimitrakopoulos05}, a generate signal $g_{i:0}$ is equal to the carry signal $c_{i}$.
Thus, a valid prefix circuit should contain prefix nodes with all possibles ranges of bits, starting from $0$.
We denote such prefix nodes as \textit{output nodes}.
In Fig.~\ref{fig:preliminary-4-bit-prefix-circuit}, it has $4$ output nodes $\{0,4,6,7\}$, marked in red.

\subsection{LLM for Structured Text Generation}
\label{subsec:LLM-Text-Generation}
\textit{Structured text} is organized according to a predefined format, syntax, or set of rules, ensuring it is interpretable and processable by humans and machines. 
Examples  include tables, configuration files, JSON, XML, and domain-specific formats. 
The generation of structured text is challenging due to the strict requirements for syntactic correctness, semantic consistency, and adherence to domain-specific constraints.

LLMs, such as \textit{Generative Pre-trained Transformer} (\textit{GPT}), have demonstrated remarkable capabilities in generating coherent and contextually relevant text across diverse applications. 
However, generating structured text presents unique challenges. 
Unlike freeform text, structured text demands precise formatting and logical consistency, making it prone to errors such as incomplete structures, misplaced elements, or logical contradictions. 
Addressing these challenges requires techniques such as \textit{prompt engineering} and \textit{iterative frameworks with refinement}, where the output is progressively improved through validation and feedback.

Iterative frameworks have emerged as a critical solution for structured text generation. 
By incorporating feedback loops, these frameworks validate the correctness of the generated text and iteratively refine it to meet the desired requirements. 
For instance, iterative mechanisms have been applied in translation refinement~\cite{chen24} and data cleaning~\cite{Ni24}, where the output quality improves with each iteration. 
The iterative process not only improves accuracy but also reduces the need for domain-specific expertise, enabling non-experts to achieve high-quality structured outputs. 
By combining the flexibility, scalability, and automation capabilities of LLMs with systematic validation and feedback mechanisms, iterative frameworks have proven to be effective in addressing the complexities of structured text generation across diverse domains.

\section{LLM-aided Prefix Circuit Design} 
\label{sec:LLM-aided-Prefix-Circuit-Design} 
In this section, we will introduce our LLM-based framework for synthesizing a valid prefix circuit with optimized area and delay.
In Section~\ref{subsec:valid_prefix_circuit}, we propose a structured text-based representation of the prefix circuit called \textit{Structured Prefix Circuit Representation (SPCR) } and transform the valid prefix circuit synthesis task into the SPCR. An iterative framework is proposed in Section~\ref{subsec:valid_prefix_circuit}  to generate a SPCR.
In Section~\ref{subsec:Prefix-Circuit-Optimization}, we propose an iterative design space exploration framework to  optimize the area and delay of the prefix circuit, i.e., iteratively synthesizing better prefix circuits.

\subsection{LLM-aided Valid Prefix Circuit Synthesis}
\label{subsec:valid_prefix_circuit}
\begin{figure}
    \centering
    \includegraphics[width=\linewidth]{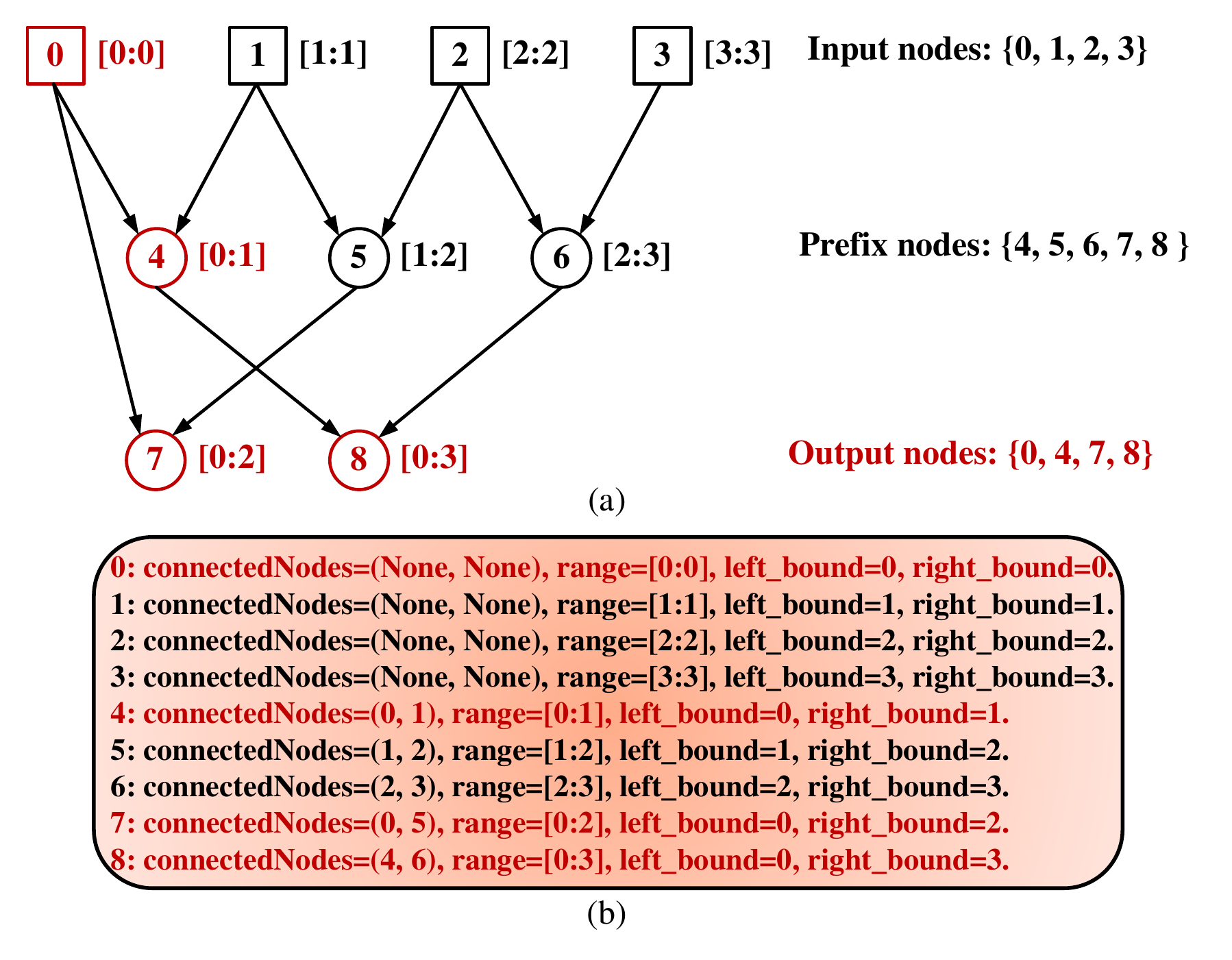}
    \caption{(a) An example 4-bit valid prefix circuit with 4 input nodes and 5 prefix nodes; (b) the corresponding SPCR.}
    \label{fig:example_prefix_circuit}
\end{figure}

In this section, we will introduce our proposed iterative framework that uses LLM to synthesize valid prefix circuits.
Our main idea is to transform the valid prefix circuit synthesis as a structured text generation task based on LLM.
We propose a structured text representation of the prefix circuit, called \textit{Structured Prefix Circuit Representation} (\textit{SPCR}) as follows:
\begin{Definition}
The Structured Prefix Circuit Representation is a standardized text-based format for representing the nodes, connections, and computational ranges in a prefix circuit. Each line in the SPCR format corresponds to a single node in the circuit and contains the following elements:
\begin{itemize}
[noitemsep,nolistsep,leftmargin=*]
    \item \textbf{Node Index}: A unique index for the node, denoted as an integer, e.g., $0$, $1$, $2$, etc.
    \item \textbf{connectedNodes}: Specifies the two predecessor nodes that provide inputs to this node. The format is (left\_node, right\_node), where: left\_node and right\_node are indexes of the two nodes. Specially, for input nodes, this field is (None, None) as they have no predecessor nodes.
    \item \textbf{range}: The bit range [left\_bound:right\_bound] that the node computes propagate and generate signals for.
    \item \textbf{left\_bound}: Start bit index of the range the node computes.
    \item \textbf{right\_bound}: Ending bit index of the range node computes.
\end{itemize}
\end{Definition}
In Fig.~\ref{fig:example_prefix_circuit} (a), it shows an example valid 4-bit prefix circuit.
The prefix circuit has 4 input nodes (nodes $0$-$3$) and 5 prefix nodes (nodes $4$-$8$). 
The bit range of each node is marked next to it.
Among them, nodes $0$, $4$, $7$, and $8$ are output nodes, which are marked in red.
The corresponding SPCR is shown in Fig.~\ref{fig:example_prefix_circuit} (b), where each line corresponds to a node in the prefix circuit and the lines marked in red are output nodes.
For example, node $5$ has two predecessor nodes $1$ and $2$ denoted as connectedNodes$=(1, 2)$, and its bit range is denoted as range$=[1:2]$ with left\_bound$=1$ and right\_bound$=2$.
For simplicity, we denote a SPCR, whose corresponding prefix circuit is valid, as a \textit{valid SPCR}.

\begin{figure}
    \centering
    \includegraphics[width=\linewidth]{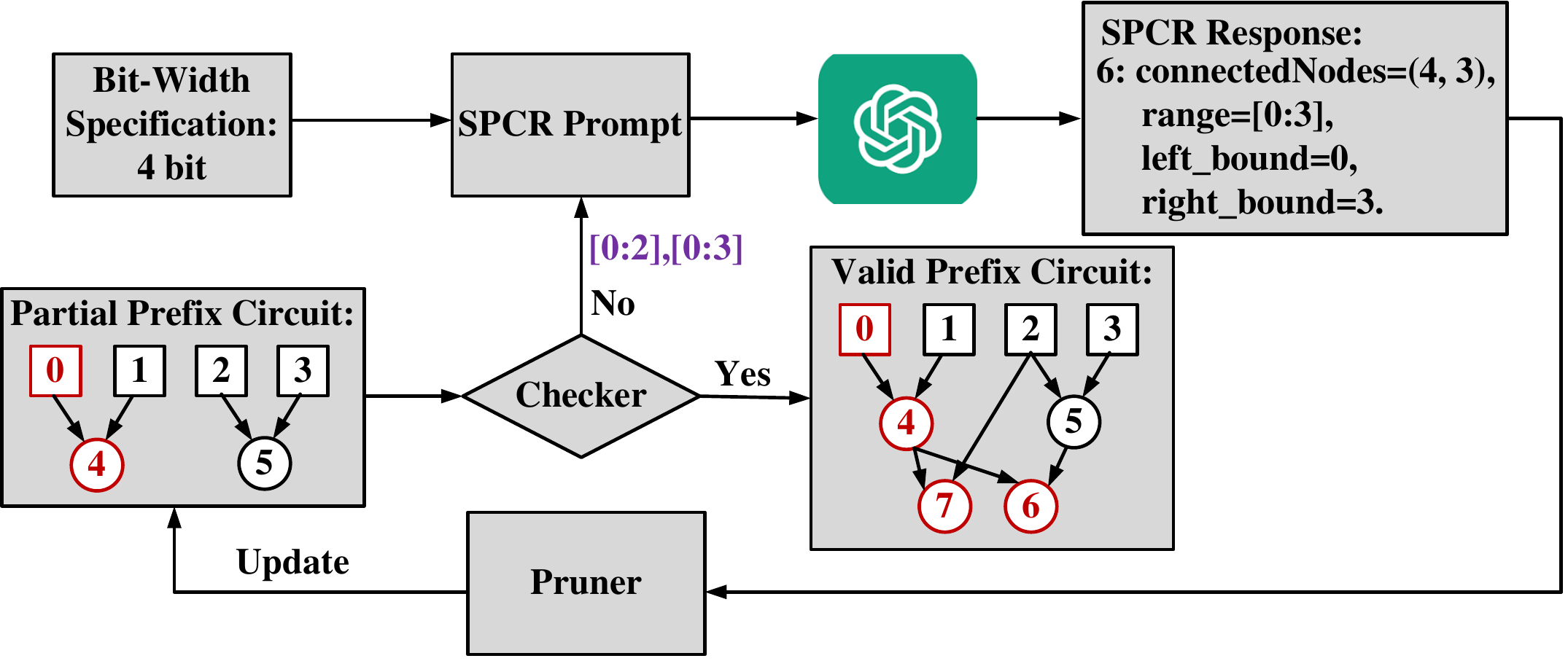}
    \caption{Iterative framework for prefix circuit synthesis.}
    \label{fig:flow-SPCR-Synthesis}
\end{figure}

\begin{figure}
    \centering
    \includegraphics[width=\linewidth]{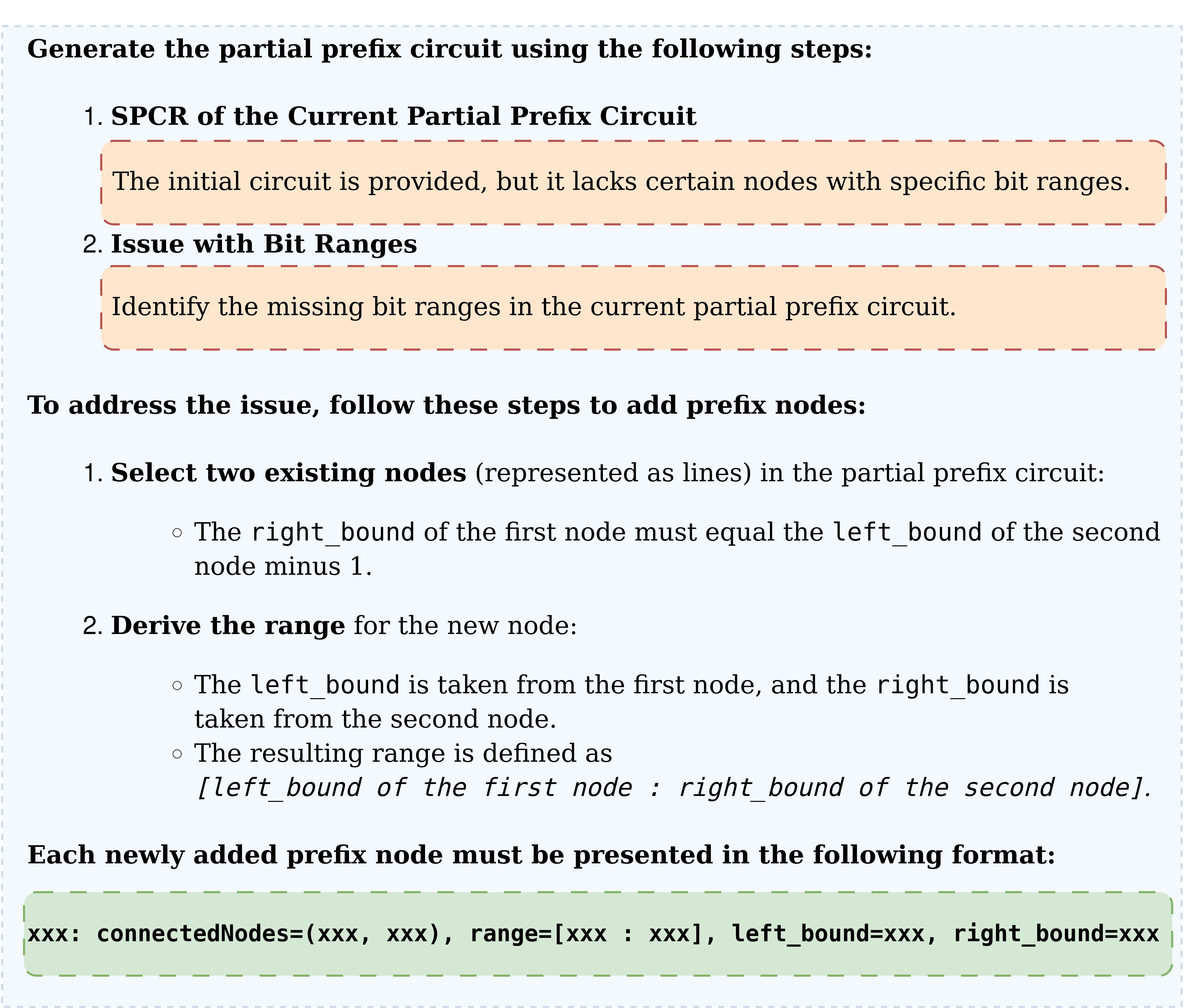}
    \caption{SPCR Prompt:  the two orange frames have to be changed according to the current partial prefix circuit, and the remaining parts are  fixed.}
    \label{fig:SPCR_Prompt}
\end{figure}

A prefix circuit has a one-to-one relationship with a corresponding SPCR. This relationship allows us to transform the valid prefix circuit synthesis problem into the  LLM-aided valid SPCR generation problem. 
We generate valid SPCRs by using an LLM, which are then used to construct valid prefix circuits. 
As is described in Section~\ref{subsec:LLM-Text-Generation}, generating the SPCR using LLM can be prone to errors.
We propose an iterative framework for generating the valid SPCR based on LLM, which is finally converted into the corresponding valid prefix circuit.
Fig.~\ref{fig:flow-SPCR-Synthesis} shows the overall flow of the iterative framework to synthesize valid prefix circuit. Steps in the flow are:
\begin{itemize}
    \item \textbf{Bit-Width Specification}: The framework begins with the specification of the bit-width $n$ for the prefix circuit (e.g., $4$-bit, $8$-bit). This specification also determines that the valid prefix circuit should contain all $n$ output nodes.
    \item \textbf{Partial Prefix Circuit}: The framework iteratively adds  valid prefix nodes into the partial prefix circuit, until all $n$ output nodes are generated. The partial prefix circuit is initialized with only $n$ input nodes in this framework.
    \item \textbf{Checker}: validates whether the partial prefix circuit in the current iteration is valid. If it is valid, the framework terminates and output this valid prefix circuit; otherwise, it will output the lacked bit ranges of the current partial prefix circuit. In Fig.~\ref{fig:flow-SPCR-Synthesis}, the checker can determine that the partial prefix circuit in the current iteration is not valid and output the lacking bit ranges $[0:2]$, $[0:3]$.
    \item \textbf{SPCR Prompt}: Given the current partial prefix circuit and the lacking bit ranges, this step will generate a prompt, called \textit{SPCR prompt}, with the format shown in Fig.~\ref{fig:SPCR_Prompt}. The initial part of the SPCR prompt contains the SPCR of the currrent partial prefix circuit and its lacked bit ranges. The next part is to ask the LLM for adding valid prefix nodes following two steps: the first is to ensure satisfying Eq.~\ref{eqn:condition_prefix_node} and the other is to derive the corresponding bit range. 
    \item \textbf{SPCR Response}: The SPCR response is the response to the SPCR prompt from the LLM, in which each line corresponds to the SPCR of a prefix node to be added. In Fig.~\ref{fig:flow-SPCR-Synthesis}, it shows an example SPCR response for the current partial prefix circuit, which represents a new prefix node $6$ with two predecessor nodes $4$ and $3$, and bit range $[0:3]$.
    \item \textbf{Pruner}: The generated new prefix nodes, however, may be invalid. The pruner is used to prune all invalid prefix nodes from the SPCR response, which does not satisfy Eq.~\ref{eqn:condition_prefix_node}. Node $6$ in the example SPCR response in Fig.~\ref{fig:flow-SPCR-Synthesis} is invalid, such that it will be pruned by the pruner. After pruning, the valid prefix nodes in the SPCR response will be added into the current partial prefix circuit.
    
\end{itemize}

\subsection{LLM-aided Design Space Exploration of Prefix Circuits}\label{subsec:Prefix-Circuit-Optimization}

\begin{figure}
    \centering
    \includegraphics[width=\linewidth]{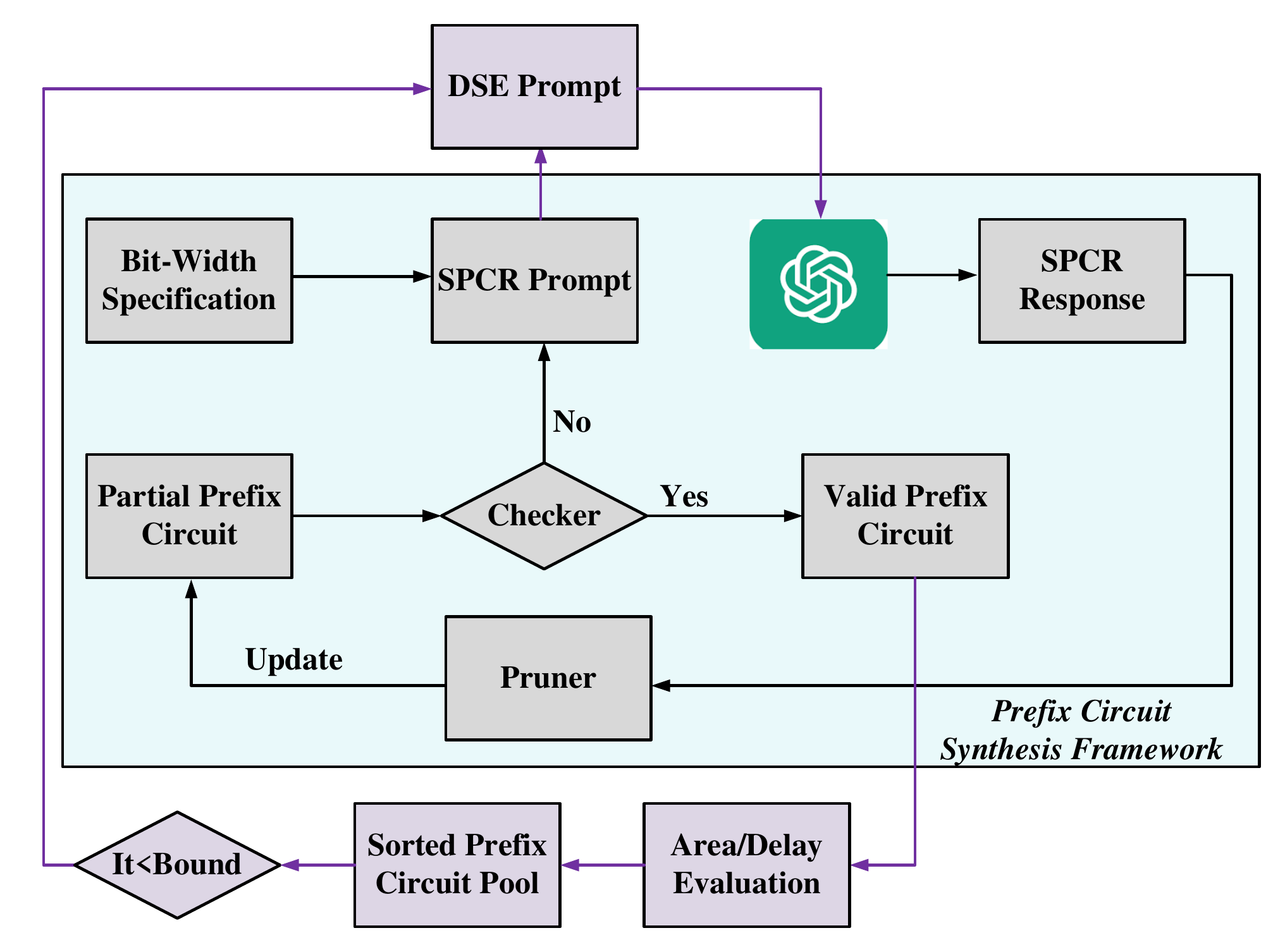}
    \caption{Iterative DSE framework for optimizing prefix circuits.}
    \label{fig:DSE-Framework}
\end{figure}

\begin{figure}
    \centering
    \includegraphics[width=\linewidth]{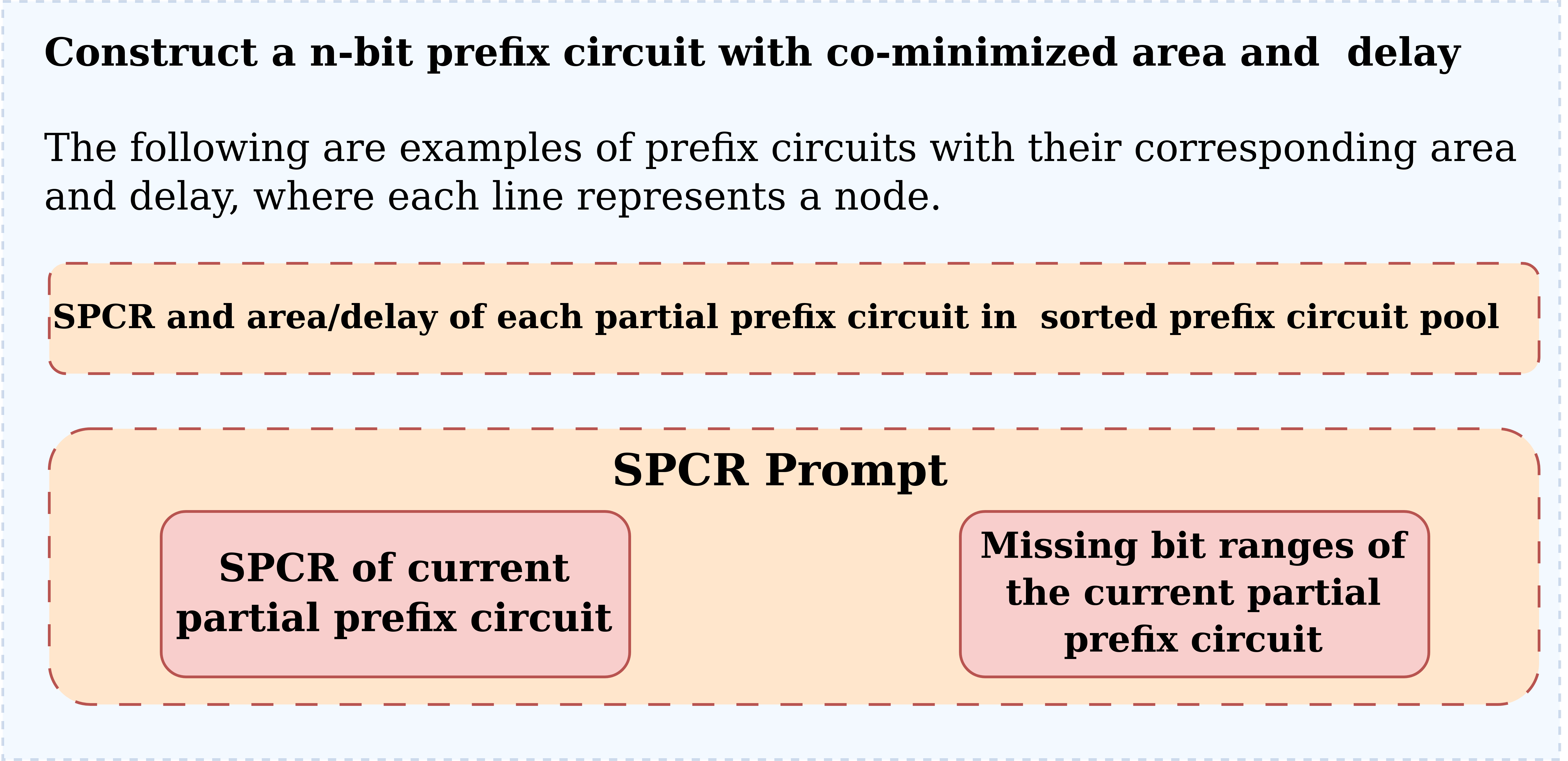}
    \caption{DSE Prompt: the SPCR and the corresponding area and delay of each prefix circuit in the sorted prefix circuit pool; the SPCR prompt of the current partial prefix circuit; the remaining parts are a fixed template.}
    \label{fig:DSE-Prompt}
\end{figure}

\begin{figure}
    \centering
    \includegraphics[width=\linewidth]{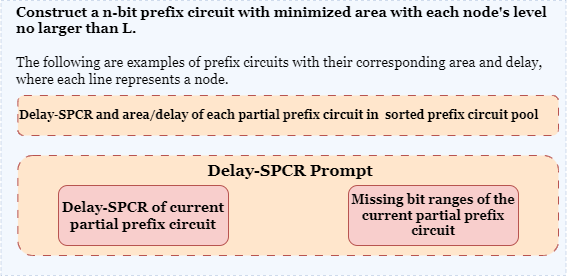}
    \caption{Modifed DSE prompt for the delay-limited area minimization mode.}
    \label{fig:modifed-DSE-prompt}
\end{figure}

In this section, we will introduce an iterative DSE framework to optimize the area and delay of the prefix circuit. Building upon the prefix circuit synthesis framework in Section~\ref{subsec:valid_prefix_circuit}, the iterative DSE framework  optimize the area and delay of prefix circuits. 
This framework incorporates a novel mechanism to use the \textit{Sorted Prefix Circuit Pool} and the \textit{DSE Prompt} for guiding the LLM toward discovering high-quality designs. 
The framework (see  Fig.~\ref{fig:DSE-Framework}) operates as follows:
\begin{itemize}
[noitemsep,nolistsep,leftmargin=*]
    \item \textbf{Area and Delay Evaluation}: Each valid prefix circuit is evaluated for area (as the number of nodes including input and prefix nodes) and delay (logic levels) Our DSE framework can  incorporate other area and delay metrics from commercial EDA tools. These metrics quantify the trade-offs between resource efficiency and computational speed and are the basis for ranking the circuits.
    \item \textbf{Sorted Prefix Circuit Pool}: All prefix circuits are stored in a sorted prefix circuit pool, ranked in the descending order of area and delay. We apply the non-dominated sorting algorithm~\cite{Deb00} to sort prefix circuits using the two metrics. Initially, when the pool is empty, we initialize it using classical prefix circuits, such as Kogge-Stone~\cite{Kogge73}.
    \item \textbf{DSE Prompt}: As shown in Fig.~\ref{fig:DSE-Prompt}, it combines the information from the sorted prefix circuit pool with the SPCR prompt of the current partial prefix circuit.
    \item \textbf{Prefix Circuit Synthesis Framework}: The core of the DSE framework is the prefix circuit synthesis framework. However, there is some difference for generating each SPCR response. Instead of querying the LLM using the SPCR prompt, the DSE framework uses the new DSE prompt to generate the SPCR response iteratively. This  provides the LLM with prior prefix circuits and  performance metrics and  allows the LLM to detect patterns and trends from prior circuits, thereby improving its ability to generate better SPCR responses for the current partial prefix circuit with greater potential for minimizing area and delay.
    \item \textbf{Iteration Bound}: is the number of iterations of the DSE framework to limit the overall runtime.
\end{itemize}
Our DSE framework can support more optimization modes that enable  a broader range of application-specific requirements, offering flexibility in DSE:
\begin{itemize}
[noitemsep,nolistsep,leftmargin=*]
    \item \textbf{Delay-Limited Area Minimization}: Minimizes  prefix circuit's area under a given delay  bound.
    \item  \textbf{Area-Limited Delay Minimization}: Targets minimizing the prefix circuit's delay, while ensuring that the area remains within a upper bound.
\end{itemize}
For brevity, we describe the delay-limited area minimization mode here and the other mode is similar to be derived.
To ensure that the synthesized prefix circuit meets a specified delay bound $L$, it is equivalent to limiting each newly added prefix node so that its delay does not exceed the given bound.
Thus, our DSE prompt is modified as shown in Fig.~\ref{fig:modifed-DSE-prompt}.
The key change is over SPCR, called \textit{Delay-SPCR}, in which each line should also include the delay of its corresponding node.
Accordingly, the LLM is queried to compute the delay of the new prefix node to be added, which is realized by adding such a prompt into our SPCR prompt: \textit{derive the level as the larger level of the two selected nodes}.
The new SPCR is called \textit{Delay-SPCR Prompt}.


\section{Experimental Results} \label{sec:exp}

\subsection{Experimental Setup} 
We implemented PrefixLLM, using Python. 
The framework leverages the \textit{OpenAI o1 mini} model, accessed via the Python \textit{OpenAI API}, as the core LLM. 
The OpenAI o1 model was selected for its robust logical reasoning capabilities. 
In the DSE framework, we set the maximum number of iterations to $20$ to balance computational efficiency and exploration. To control the length of the DSE prompt, we included  the $10$ best prefix circuits from the sorted prefix circuit pool, which were ranked based on their area and delay. In this section, we evaluated PrefixLLM in the depth-limited area minimization mode, where the objective is to minimize area given a predefined depth bound. 

Experiments were conducted on $8$-bit and $16$-bit prefix circuits, and the results were compared with the state-of-the-art machine learning-based methods~\cite{Lai24}, and three classical prefix circuits Sklansky~\cite{Sklansky60}, Kogge-Stone~\cite{Kogge73} and Brent-Kung~\cite{Brent82}.
Note that the work in~\cite{Lai24} is the most state-of-the-art open-source machine learning-based method, which can outperform the work in~\cite{Roy13} and PrefixRL~\cite{Roy21}.
Both our PrefixLLM and the work in~\cite{Lai24} need a initial prefix circuit to start the design space exploration, which is set as Kogge-Stone in this section.
The Kogge-Stone is known for its minimal delay but incurs significant area overhead due to its exhaustive structure, the Brent-Kung offers a trade-off between area and delay with more compact designs at the cost of slightly larger delay, and the Sklansky has the minimal delay as well and smaller area than Kogge-Stone but often results in increased fanouts.

Finally, we use the \textit{Synopsys Design Compiler}~\cite{DC} to measure the actual area and delay of the synthesized prefix circuits using the \textit{Nangate 45nm Technology}~\cite{nangate}.

\subsection{Comparison of Synthesized Prefix Circuits}
\label{subsec:comparison}
Table~\ref{tab:size_comparison}   compares the area (number of nodes) for various prefix circuits under different input bit widths and delay bounds (number of logic levels). 
The evaluated circuits include Sklansky, Kogge-Stone, Brent-Kung, the state-of-the-art machine learning-based method~\cite{Lai24}, and the proposed PrefixLLM framework. 
The theoretical area bounds are also shown. 
To highlight the most efficient designs, the smallest area for each delay bound and bit-width is marked in bold.

For $8$-bit prefix circuits, PrefixLLM synthesizes the best prefix circuit for each delay bound. 
At delay bound $4$, PrefixLLM achieves the best prefix circuit with $20$ nodes, with a reduction of area by $2$ compared with~\cite{Lai24}. 
Across all delay bounds, PrefixLLM achieves a total size of $86$ nodes with a $2.27$\% reduction compared to the state-of-the-art method~\cite{Lai24}. 
Additionally, PrefixLLM can outperform all of the three classical prefix circuits under the same delay.

For $16$-bit circuits, PrefixLLM shows improvements over  baselines. Across all delays other than $6$, PrefixLLM achieves a size of $208$ nodes, representing a $3.70$\% reduction compared to the $216$ nodes of~\cite{Lai24}. 
At delay bound $5$, PrefixLLM synthesizes a prefix circuit with $47$ nodes, outperforming Sklansky ($48$ nodes) and the state-of-the-art  ($58$ nodes). 
At delay bounds $7$, $8$, and $9$, PrefixLLM synthesizes circuits with the same areas as those generated by~\cite{Lai24}, close to the theoretical bound. 
Thus
PrefixLLM balances area and delay efficiently, even for higher bit-width circuits.

Table~\ref{tab:size_comparison} shows that PrefixLLM  minimizes the area of prefix circuits for different delay (depth) bounds. Under strict delay bounds such as $4$ for $8$-bit circuits and $5$ for $16$-bit circuits, PrefixLLM outperforms~\cite{Lai24}.
\begin{table}[!thbp]
\centering
\tabcolsep = 3pt
\caption{Comparisons of prefix circuits in area and delay.}
\begin{tabular}{c c|c|c c c c c}
\toprule\textbf{\tabincell{c}{Input\\Bit}} & \textbf{Delay} & \textbf{\tabincell{c}{Theory\\Area\\Bound\\ \cite{Snir86}}}  & \textbf{Sklansky} & \textbf{\tabincell{c}{Kogge\\Stone}} & \textbf{\tabincell{c}{Brent\\Kung}} & \textbf{\cite{Lai24}} & \textbf{PrefixLLM} \\
\midrule
8 & 4 & 18 & \textbf{20} & 25 & - & 22 & \textbf{20} \\
8 & 5 & 17 & - & - & 19 & \textbf{18} & \textbf{18} \\
8 & 6 & 16 & - & - & - & \textbf{17} & \textbf{17} \\
8 & 7 & 15 & - & - & - & \textbf{16} & \textbf{16} \\
8 & 8 & 14 & - & - & - & \textbf{15} & \textbf{15} \\
 &    &    &   &   &   &      88    &  \textbf{86 (2.27\% $\downarrow$)} \\
\midrule
16 & 5 & 41 & 48 & 65 & - & 58 & \textbf{47} \\
16 & 6 & 40 & - & - & - & \textbf{41} & 44 \\
16 & 7 & 39 & - & - & 42 & \textbf{40} & \textbf{40} \\
16 & 8 & 38 & - & - & - & \textbf{39} & \textbf{39} \\
16 & 9 & 37 & - & - & - & \textbf{38} & \textbf{38} \\
&    &    &   &   &   &      216    &  \textbf{208 (3.70\% $\downarrow$)} \\
\bottomrule
\end{tabular}
\label{tab:size_comparison}
\end{table}

\begin{figure}[bthp]
    \centering
    \includegraphics[width=\linewidth]{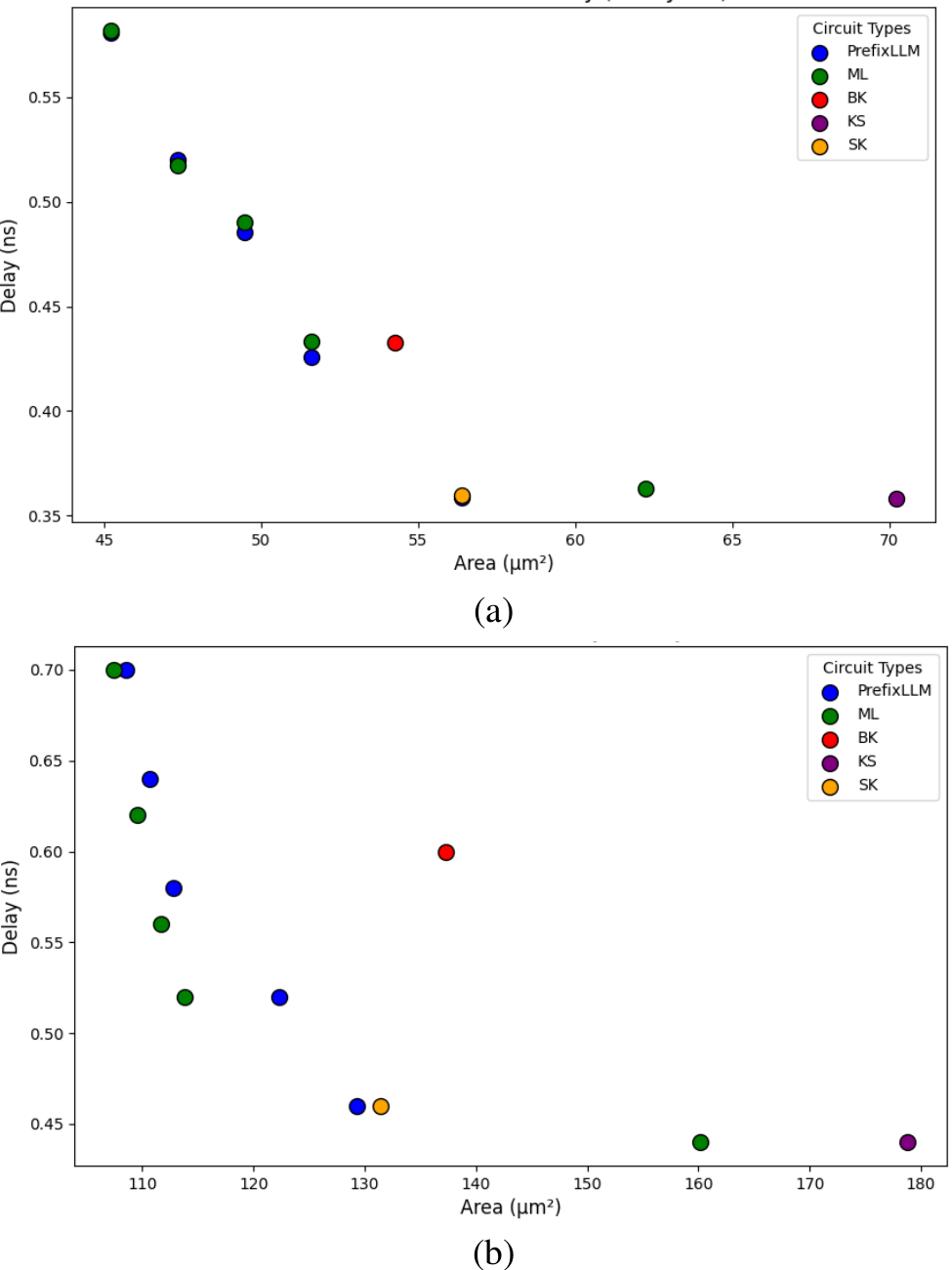}
    \caption{Evaluation of acutal area ($\mu m^2$) and delay ($ns$) of prefix circuit-based adders: (a) $8$-bit; (b) $16$-bit.}
    \label{fig:real-area-delay}
\end{figure}

To  validate the effectiveness of  PrefixLLM, we use these prefix circuits to construct $8$- and $16$-bit digital adders and evaluate their  area and delay by Synopsys Design Compiler using Nangate 45nm technology.
The comparison under $8$- and $16$-bit cases are shown in Fig.~\ref{fig:real-area-delay} (a) and (b), respectively. 
\textit{PrefixLLM} corresponds to our proposed work, \textit{ML} corresponds to the machine learning-based work~\cite{Lai24}, \textit{BK} corresponds to Brent-Kung, \textit{KS} corresponds to Kogge-Stone, and \textit{SK} corresponds to Sklansky.
From the results,  prefix circuits by PrefixLLM outperform classical prefix circuits for $8$- and $16$-bit adders.
Compared with~\cite{Lai24} for  $8$-bit adders, designs by  PrefixLLM achieve the similar performance over area and delay, except the case under the delay bound $4$ that the adder based on our prefix circuit can achieve smaller area.
For $16$-bit adders, our prefix circuits achieve the similar area and delay compared with those using prefix circuits from~\cite{Lai24}, except the case of delay bound $5$.
The adder based on our prefix circuit achieves  more reduction in area compared to that using prefix circuit from~\cite{Lai24} for the same delay bound.

\subsection{Chip Tapeout} 
To show the practicality and robustness of the prefix circuits, 
We took an $8$-bit adder using the prefix circuit generated by our PrefixLLM and went through the Tiny Tapeout process~\cite{Venn24}.
The Tiny Tapeout process integrates OpenLane, which is an automated RTL to GDSII flow, and is based on the \textit{SkyWater} $130$nm open source technology.
In the process, the designs are implemented on each Tiny Tapeout tile, which is about $160\times 100 \mu m^2$.

\begin{figure}
    \centering
    \includegraphics[width=\linewidth]{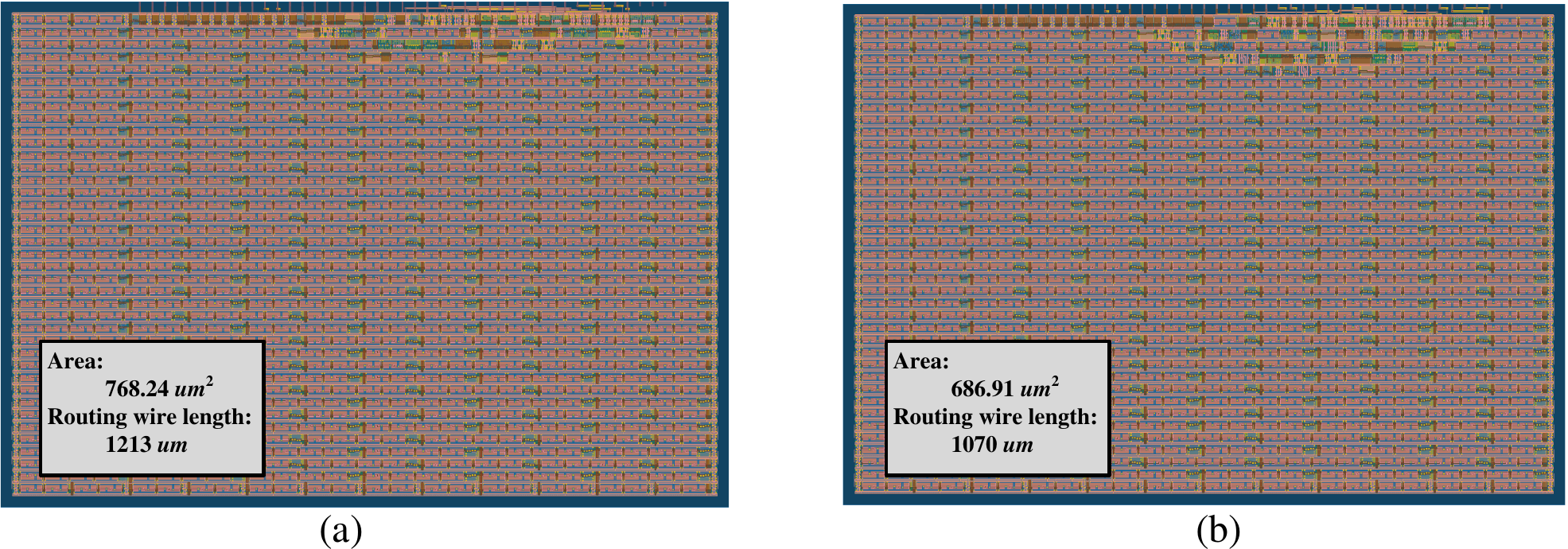}
    \caption{GDS of two $8$-bit prefix circuit-adders based on (a) Kogge-Stone prefix circuit and (b) prefix circuit generated by PrefixLLM.}
    \label{fig:tiny-tapeout}
\end{figure}

In Fig.~\ref{fig:tiny-tapeout} (a) and (b), they show two generated GDS of a single Tiny Tapeout tile, on which two $8$-bit adders are implemented using the Kogge-Stone prefix circuit and our PrefixLLM prefix circuit, respectively. 
In the lower left corner, the corresponding area and routing wire length of the two adders are displayed.
From the result, we can see that the area of the adder based on our PrefixLLM prefix circuit ($686.91\mu m^2$) achieves a $10.59\%$ reduction over that of the Kogge-stone based adder ($768.24\mu m^2$).
Moreover, the routing wire length is another important metric, which has great effects over digital circuits' delay and power consumption.
From Fig.~\ref{fig:tiny-tapeout}, the routing wire length of the adder based on our PrefixLLM prefix circuit ($1070\mu m$) is $11.79\%$ smaller than that of the Kogge-stone based adder ($1213\mu m$).

By leveraging the tiny tapeout process, we demonstrate the applicability of our PrefixLLM framework in generating prefix circuits that can seamlessly transition from theoretical synthesis to physical hardware realization.

\section{Conclusion} \label{sec:con}
By introducing the SPCR, we transformed the valid prefix circuit synthesis into a structured text generation problem, enabling efficient and automated prefix circuit synthesis using LLM. 
However, SPCR generation is inherently error-prone due to the strict structural and computational constraints of prefix circuits. 
To address this challenge, we proposed an iterative framework that ensures the automatic and correct generation of valid SPCRs using LLMs.
Building on the iterative framework for valid SPCR generation, we developed an iterative DSE framework aimed at optimizing area and delay of prefix circuits. 
By utilizing the proposed sorted prefix circuit pool and the innovative DSE prompt, the framework guides the LLM to detect useful patterns from prior designs, facilitating the discovery of optimized designs.

\bibliographystyle{IEEEtran}
\bibliography{Reference}

\end{document}